\documentstyle[epsf,onecolumn]{mn}
\title{A mini-supernova model for optical afterglows of GRB}
\author[S.I.Blinnikov and K.A.Postnov] 
{S.~I.Blinnikov$^1$ and K.A.Postnov$^2$\\
$^1$ Institute for Theoretical and Experimental Physics, 
117259 Moscow, Russia\\
$^2$ Faculty of Physics, Moscow State University, 119899 Moscow, Russia}

\date{Accepted 1997 ...
      Received 1997 ...}
\pagerange{\pageref{firstpage}--\pageref{lastpage}}
\pubyear{1997}
\begin{document}
\maketitle

\label{firstpage}

\begin{abstract}
An energy deposition of $\sim 10^{50}$ ergs into the exterior $10^{-3}$
M$_\odot$ layers of a red giant is calculated to produce an optical
phenomenon similar to afterglows of gamma-ray bursts (GRB) recently
observed.  This mechanism can be realized if a GRB is generated by
some mechanism in a close binary system. In contrast to a ``hypernova''
scenario for GRB recently proposed by Paczy\'nski (1997), this model
does not require a huge kinetic energy of the expanding shell to explain
optical afterglows of GRB.

\end{abstract}

\section{Introduction}

Recent multiwavelength identification of two BeppoSAX GRB 970228 (van
Paradijs et al. 1997; Galama et al. 1997) and GRB 970508 (Piro et
al. 1997) became a major astronomical ``event of the year''.  HST
observations of a possible ``host galaxy'' around a point-like optical
transient to GRB 970228 and measurements of a high redshift $z=0.835$
of the absorption and emission lines in the spectra of optical
counterpart to GRB 970508 (Metzger et al. 1997) have provided a strong
evidence for the cosmological origin of GRB.  If so, the energy
deposited in gamma-rays during GRB should be $\sim 10^{51}$ ergs.  A
power-law brightness decline with time observed in both optical
afterglows $\sim t^{-1.1}$ (Galama et al. 1997, Kopylov et al. 1997)
has readily been explained by the standard model of a relativistic
fireball interacting with adjacent material, either interstellar or
intergalactic (Rees and M\'esz\'aros 1992; e.g. Wijers et
al. 1997, Waxman 1997, and references therein).

However, the apparent success of the relativistic fireball model in
explaining the observed optical afterglows of GRB still uses some
{\em a priori} assumptions about the ultimate source of the fireball itself. A
critical consideration of the fireball formation under real
astrophysical conditions has recently motivated Paczy\'nski (1997) to
suggest an alternative model for GRB, which he calls ``a hypernova
model'' and which involves a huge energy deposit ($\sim 10^{54}$
ergs) into the kinetic energy of the expanding envelope of a type Ib
supernova (i.e., originating from the core collapse of a helium star)
in a close binary system via extremely strong ($\sim 10^{15}$ G)
magnetic field of the nascent rapidly rotating neutron star.

Originally, the idea of magneto-rotational supernova explosion was
proposed by Bisnovatyi-Kogan (1970) and further elaborated during
three decades, see Ardeljan et al. (1996ab, 1997). A very strong
magnetic field of a rapidly rotating neutron star as a source of GRB
was proposed by Usov (1992). The possibility of a GRB to appear during
a bare core collapse in a binary system was suggested by Dar et
al. (1992).  The latter model assumed a GRB to be a result of the
neutrino-antineutrino pair creation and annihilation during the
accretion-induced collapse of a white dwarf in a close binary system,
when only a tiny fraction of mass is ejected ($< 10^{-4}$ M$_\odot$).

In this paper, we suggest another conceivable model for optical
afterglows to GRB, which involves a mild energy release of the order
$10^{50}-10^{51}$ ergs in gamma-rays in a close binary system. The
ultimate source of these gamma-rays may be associated with bare
collapse of an accreting white dwarf, perhaps with some unknown
mechanism (e.g., exotic particle decays).  In principle, this
mechanism can be of the same nature as the one leading to a standard
supernova type II explosion (when the collapse occurs inside a massive
star). However, the precise nature of the GRB is not important for the
purpose of the present letter; we wish focus only on the consequences
the gamma-ray irradiation would have in a close binary system.  In
contrast to Paszy\'nski's hypernova scenario, we need no a huge
kinetic energy of the expanding envelope to produce a prolonged
optical afterglow. We show that deposition of the energy of gamma-ray
burst into the exterior envelope of a red supergiant in a close binary
system could be sufficient to explain optical afterglow decaying with
time in a way similar to that observed in GRB 970228.

\section{The model}

Consider the following model.  Let a GRB phenomenon with the energy of
the order of a SN explosion occur in a close binary system, e.g.  as a
result of accretion-induced collapse of a WD in a symbiotic binary
system. This process is currently one of the favoured mechanisms for
SN type Ia explosions (Nomoto et al. 1997).  In our model we do not
specify the detailed mechanism of the GRB; for example, it can be
associated with a `failed' SN (Woosley 1993) or something else.  Let
the secondary red giant companion be close enough to intercept a
significant fraction of gamma-rays emitted.  Let pose the question:
What is the consequence of such immediate gamma-ray irradiation of the
red giant atmosphere? For simplicity, we consider a spherically
symmetric energy deposition in the red giant atmosphere. To calculate
the afterglow, we used a multi-energy group radiation-hydro-code
STELLA (Blinnikov \& Bartunov 1993; Blinnikov et al. 1997). The method
uses the time-dependent equations for the angular moments of intensity
averaged over fixed frequency bands and solves the non-relativistic
equations of hydrodynamics implicitly coupled to them.

We consider a red supergiant with a mass of 15 $M_\odot$ and radius
$r\sim 4000 R_\odot$. Actually, the mass of the star is of minor
importance since the energy was deposited only into the exterior
layers with a mass of $10^{-3} M_\odot$.
The deposition of $1.5\times 10^{50}$ ergs of gamma-ray energy into
these layers during 10 seconds, assumed in our model calculation,
heats them up to $\sim 5\times 10^5$ K. The thermal energy thus stored
is orders of magnitude higher than the initial thermal energy of the
exterior layers of the red giant star. The energy budget is spent
twofold: a part of it is transferred as a heat wave into deeper layers
(down to $\sim 0.1 M_\odot$ during $\sim 0.1$ day) and another part
goes into the thermal and kinetic energy of the outermost layers. All
layers out of $10^{-4}M_\odot$ acquire speed of $\sim 2\times 10^4$
km/s during a few hours, while the layers interior to $10^{-3}M_\odot$
remain almost at rest with the speed less than $10^3$ km/s. By this
time the temperature of the outer layers falls down by an order of
magnitude, to $\sim 5\times 10^4$ K.  The inward moving heat wave dies
out during $\sim 10$ days reaching around $ 1 M_\odot$. By that time
the kinetic energy of the expanding shell is about 2\% of the initial
energy deposited into envelope.
The resulting light curve is shown in Fig. 1. We assumed a 1 Gpc
distance to the source. No redshift corrections were made, so in fact
the apparent R-magnitudes are close to the intrinsic V-magnitudes.

The observed magnitudes of the optical transient of GRB 970228 are
taken from Galama et al. (1997).
It is seen from Fig. 1 that the resulting light curve follows well the
observed behaviour of the optical transient associated with GRB
970228, especially during first 10 days. At later epoch, near 30 days,
the calculated optical fluxes becomes two magnitudes smaller than the
observed ones (the second HST point for the day 39, $R = 25.5$, is not
shown in Fig. 1).
However, we can say that taking into account of the mini-blast wave
interaction with circumstellar medium (pre-supernova wind) at later
stages could increase the fluxes and make the agreement with
observations better.

If more energy is deposited the effect would be more dramatic -- the
afterglow would be brighter and longer.

In the case of a GRB occurring in a classical high-mass X-ray binary
(due to, for example, collapse of a neutron star to a black hole) with an
O/B supergiant as the companion, the consequences would change
significantly.  The star being more compact, the mass of the outer
shell heated by gamma-rays is smaller, scaling as square of the radius
(e.g., for $R \approx50 R_\odot$ the heated mass is $\sim 10^{-7}
M_\odot$). The energy deposition per unit mass is correspondingly
higher, so the speed and temperature of the matter increase as well
whereas the duration of the afterglow decreases. For the case of a $R
\approx 50 R_\odot$ supergiant we obtain $T \sim 5\times 10^6$ K and $v
\sim c$. Currently, the code STELLA is unable to treat relativisitic
problems; nevertheless, it is conceivable that the prolonged
afterglows to GRBs could be explained by the interaction of the
relativistic ejecta with the surrounding ISM. In this respect our
model converges with the Rees and M\'esz\'aros (1992) relativistic
fireball model.

\section{Discussion}

Our simplified calculations of the effect of $\sim 10^{50}$ ergs
impinging the atmosphere of a nearby red giant demonstrate a fair
correspondence with the observed light curve of the optical transient
to GRB 970228.  If more energy is released in a binary system (for
example, if even 1\% of the Paczy\'nski's hypernova energy,
i.e. $10^{52}$ ergs in gamma-rays, and/or kinetic energy), an enormous
optical flash similar to a very bright unusual supernova could be
produced by interaction of gamma-rays and expanding blast wave with
the secondary companion. Impact of $10^{54}$ ergs of kinetic energy on
the secondary optical star would probably have really dramatic
consequences. However, we think that in order to produce gamma-ray
burst and optical afterglow, one needs not to assume such high
energies -- a bare collapse of an accreting white dwarf in a binary
system could produce both gamma-ray and optical emission.

What should be other effects of a $10^{51}$-erg GRB on the surrounding
medium?

The effects must be twofold: first, the impact of $10^{51}$ ergs in
gamma-rays on the surrounding medium and second, the impact of a blast
wave expanding in the dense circumstellar medium formed by a strong
stellar wind of the pre-supernova.

The first group of effects was recently addressed by Bisnovatyi-Kogan
and Timokhin (1997). They showed that a very prolonged optical
afterglow lasting for tens of years should result from a cosmological
GRB depending on conditions in interstellar or intergalactic medium
where the GRB occurred.

The second group of effects has been extensively studied in the
context of the interaction of supernova remnants with surrounding
medium.  A non-relativistic blast wave produced by supernovae has
shown to be capable of producing large superbubbles in the
interstellar medium (for example, the well known Cygnus superbubble
was suggested by Blinnikov et al. 1982 to be a result of a peculiar
supernova explosion; more recently the superbubbles are thought to be
the result of a chain of normal SN explosions, e.g. Koo, Heiles and
Reach 1992).  A superbubble would live for millions years in the dense
ISM. If occurred in a star formation region, such a hypernova would
likely to stop completely the surrounding star formation.  Huge
expanding voids in neutral hydrogen that require the energy deposition
of 100 supernovae have indeed recently been detected in several nearby
dwarf galaxies, with no apparent star formation region and associated
stellar clusters inside it (Radice et al. 1995).  This possibly may be
related to such hypernovae, but with the energy of order $\sim
10^{53}$, not $\sim 10^{54}$ ergs.

Some other difficulties that the hypernova scenario meets include (1)
a failure to transfer more than a few per cent of the rotational
energy into the kinetic energy of the envelope by the
magneto-rotational mechanism, as hydrodynamical calculations show
(Ardeljan et al. 1996ab, 1997); (2) troubles with the shock break-out
through the exploding star envelope (Blinnikov and Nadyozhin 1990;
Blinnikov et al. 1991; Ensman and Burrows 1992, and references
therein).

Therefore we propose another conceivable model for optical afterglow
of GRB, which includes a collapse of a "naked" stellar core (Ne-Mg-O,
or even iron) in a binary system as a possible source of a moderate
($\sim 10^{51}$ erg) gamma-ray burst (e.g., Dar et al. 1992).
Although it is hard to reproduce the detailed picture of GRB within
the framework of a particular model (e.g., Dar's model of the fireball
formation by neutrino-antineutrino annihilation is customarily
rejected on the grounds of being too contaminated by baryonic load,
e.g. Woosley 1993), we must note that the present situation even with
supernova explosion themselves is far from being completely clear
(e.g., Mezzacappa et al. 1997 for a recent review of difficulties with
mass ejection in supernova explosions).

Another plausible way of forming GRB at cosmological distances
involves binary neutron star merging (as originally proposed by
Blinnikov et al. 1984; see PaÓzy\'nski 1986, Lipunov et al. 1995, etc.). 
However, as detailed hydrodynamical calculations currently
demonstrate, this mechanism also fails in producing powerful fireballs
by several orders of magnitude (Janka and Ruffert 1996, Ruffert et
al. 1997).

But supernovae do explode and binary neutron stars should coalesce in
nature due to gravitational radiation regardless of the present
failure to obtain powerful fireballs in numerical
calculations. Physics is always richer than we can imagine. For
example, taking into account magnetic moment of neutrino can help in
producing envelope ejection in supernova explosions (Dar 1987;
Voloshin 1988; Blinnikov and Okun' 1988, Akhmedov et al. 1997).
If neutrino magnetic moment indeed plays the role in supernova
explosion, its interaction with outer magnetic field (the stronger the
magnetic field, the more effective the transformation of the sterile
neutrinos to `normal' ones is) in the case of an almost naked
presupernova could give rise to gamma-ray emission with temporal
features similar to the observed fractal or scale-invariant properties
found in gamma-ray light curves of GRB (Shakura et al. 1994; Stern and
Svensson 1996).  Such a temporal behaviour appears natural if
turbulence and magnetic fields are crucial in any mechanism of
gamma-ray emission (see Stern and Svensson 1996 and references
therein; see also Bykov and Toptygin 1993, Bykov and M\'esz\'aros
1996; for a review of the role of fractals, intermittency, etc.  in
the magneto-hydrodynamic turbulence see Isichenko 1992).

Electromagnetic properties of neutrino or decays of exotic particles
(Raffelt 1996) during such a dramatic process as supernova explosions
and binary compact star coalescence may directly produce the energies
required in gamma-rays ($10^{51}$ erg).  We thus may suppose that when
this energy is released at the center of a thick envelope, this leads
to the phenomenon of supernova explosion; opposedly, when the matter
coating is thinner, gamma-rays may be directly observed as a GRB, and
when this occurs in a binary system, afterglows at different
wavelengths should be produced as a consequence of interaction of
gamma-rays with secondary companion and/or local interstellar medium.

Our calculations demonstrate that this may be the case for GRB 970228.
Of course, detailed features of the light curve of the optical
transient to GRB 970228 cannot be reproduced in our simplified
spherically-symmetric calculations. The model in its simplest form
gives no explanation to the delayed optical afterglow, as observed in
GRB 970508.  One can speculate here on the role of 2D effects, on the
effects of the shock wave propagation in the circumstellar material,
etc.  The model also meets difficulties with explaining X-ray
afterglows -- but the latter can be produced either on the surface of
the 'failed' supernova remnant or connected directly with the
mini-supernova shock; some supernovae are known to emit X-rays not
connected with radioactive decay.

So other direct consequences of the $10^{51}$ erg energy deposition
into the dense pre-supernova wind, e.g. like in SN 1979C, should be
considered.  For example, taking into account of radiation transport
in the presupernova wind was shown to strongly affect the SN 1979C
light curve for many years (Bartunov and Blinnikov 1992, Blinnikov and
Bartunov 1993); Chugai and Danziger (1993) considered the interaction
of the supernova blast wave with dense clumpy wind of the precursor to
peculiar SN 1988Z to explain observed properties of its optical
emission.  As mentioned above, a very prolonged (years and tens of
years) optical afterglows may be produced by a gamma-tray burst in
interstellar medium (Timokhin and Bisnovatyi-Kogan 1995, 1996).
The latest HST observations of GRB970228 of September 5, 1997, which 
spotted the optical afterglow at $R=27.7$ (Fruchter et al. 1997) after 
half a year strongly evidence the interaction of an expanding
envelope with the surrounding medium.

We conclude that effects of supernova explosions and a powerful
gamma-ray burst on the surrounding medium and possible binary
companion may give rise to a large variety of optical
afterglows. Future observations of optical transient to GRB 970228 and
GRB 970508 will help in distinguishing these possibilities.

\section*{Acknowledgements}
The authors thank Prof. I.L.Rosental' and I.Belousova for helpful
discussions and the referee, R.A.M.J.Wijers, for his stimulating comments.
SIB acknowledges the support from the INTAS grant ``Thermonuclear
Supernovae''. KAP thanks the support from the RFBF grant 95-02-06053a.


\clearpage
\label{lastpage}
\begin{figure}
\epsfxsize=\hsize
\epsfbox{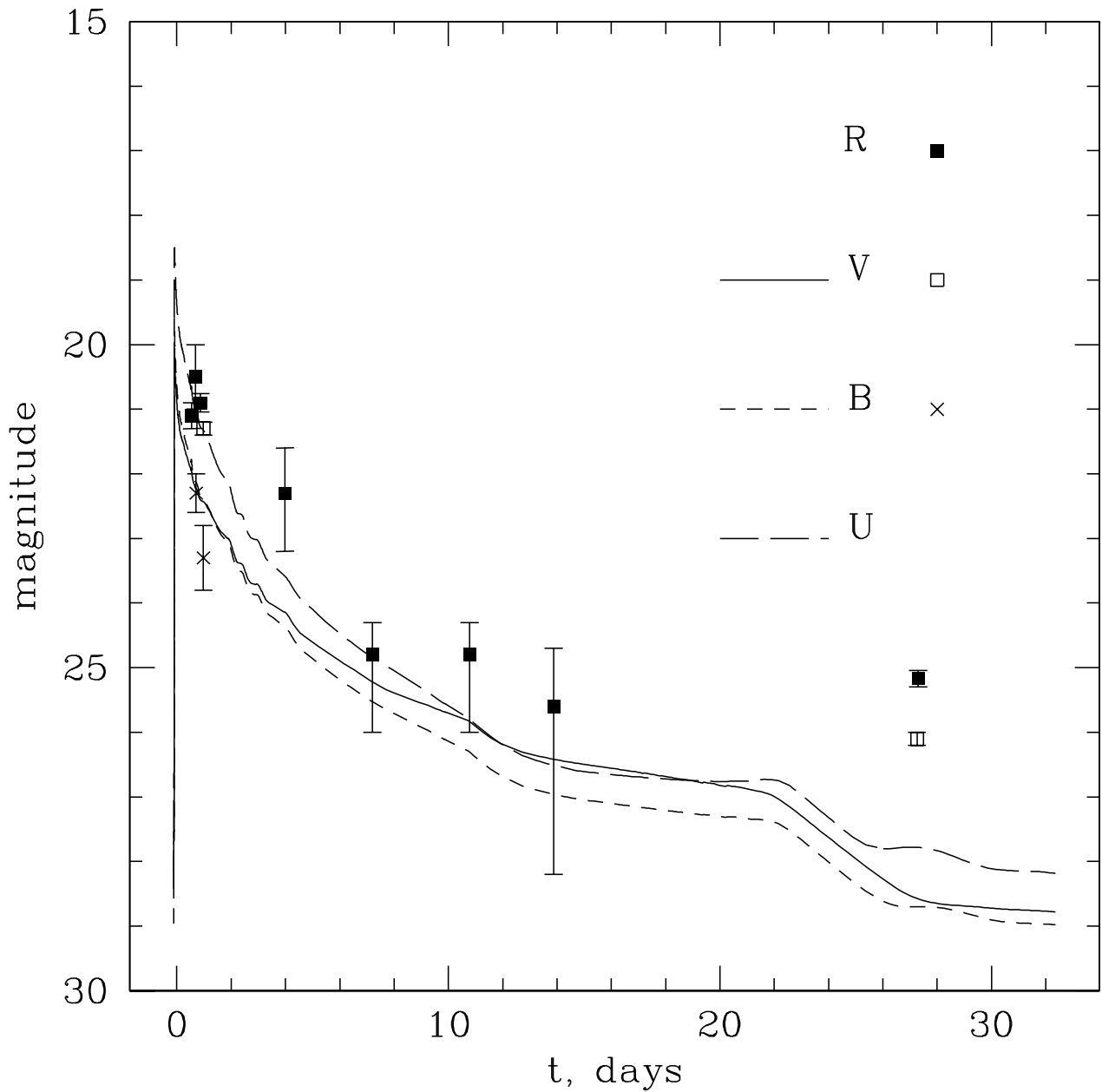}
\caption{
A mini-supernova U,B,V,R-afterglow produced by depositing $10^{50}$ ergs into
the exterior $10^{-3} M_\odot$ of a 15 $M_\odot$ red supergiant 
with a radius of $4000 R_\odot$, as seen form a distance of 1 Gpc.
The observed B,V,R-light curve of the optical transient of GRB 970228
(Galama et al. 1997) are also indicated} 
\end{figure}

\end{document}